\RequirePackage{fix-cm}
\documentclass[smallextended]{svjour3}       % onecolumn (second format)
\smartqed  % flush right qed marks, e.g. at end of proof
\usepackage{graphicx}
\begin{document}

\title{Data challenges of time domain astronomy%\thanks{Grants or other notes
}


\author{Matthew J. Graham  \and
        S. G. Djorgovski \and  Ashish Mahabal \and Ciro Donalek \and Andrew Drake \and Giuseppe Longo %etc.
}

\authorrunning{Graham et al.} % if too long for running head

\institute{Matthew J. Graham \and S. G. Djorgovski \and Ashish Mahabal \and Ciro Donalek  \and Andrew Drake\at
              California Institute of Technology,
              Pasadena, CA 91125\\
              Tel.: +1-626-395-8030\\
              Fax: +123-45-678910\\
              \email{mjg@caltech.edu}           %  \\
           \and
            \at
              Giuseppe Longo \at
              Department of Physics, University Federico II
Complesso Universitario di Monte Sant'Angelo
via Cinthia, I-80126, Naples, Italy\\
}

\date{Received: date / Accepted: date}

\maketitle

\begin{abstract}
Astronomy has been at the forefront of the development of the techniques and methodologies of data intensive science for over a decade with large sky surveys and distributed efforts such as the Virtual Observatory. However, it faces a new data deluge with the next generation of synoptic sky surveys which are opening up the time domain for discovery and exploration. This brings both new scientific opportunities and 
fresh challenges, in terms of data rates from robotic telescopes and exponential complexity in linked data, but also for data mining algorithms used in classification and decision making.
In this paper, we describe how an informatics-based approach -- part of the so-called "fourth paradigm" of scientific discovery -- is emerging to deal with these. We review our experiences with the Palomar-Quest and Catalina Real-Time Transient Sky Surveys; in particular, addressing the issue of the heterogeneity of data associated with transient astronomical events (and other sensor networks) and how to manage and analyze it.

\end{abstract}

\section{Introduction}
\label{intro}
Astronomy occupies a sweet spot in the spectrum of data-intensive sciences: the data are neither too homogeneous, originating from a single facility, like high energy physics, nor are they too heterogeneous with scores of different representations, like genetics. The data volumes are impressive, in the Tera-/Petascale regime, but not unmanageable, assuming a continuing Moore's law growth trend in hardware capabilities. Data spaces are complex and multidimensional with typically tens to hundreds of dimensions rather than thousands or millions. This makes it an ideal development and testing ground for the techniques and methodologies that are required in other data-intensive sciences.

Over the past two decades, astronomy transitioned from a relatively data-poor science to an immensely data-rich one, with the exponential growth of data volumes (and, equally importantly, data complexity and data quality) following Moore's law \cite{ref1}.  The principal agent of change were large digital sky surveys, most notably the Sloan Digital Sky Survey (SDSS; \cite{ref2}), but also DPOSS \cite{ref3}, 2MASS \cite{ref4}, and many others.

These new surveys build on a legacy of over fifty years of experience of sky surveys, first with photographic plates and then, more recently,  digital detectors (see \cite{djorgovski} for a recent review). To cope with distributed collections of giga- and terascale data collections,  the community developed the concept of the {\em Virtual Observatory} (VO; \cite{VO}). This provides the wherewithal to aggregate and analyze disparate data sets, opening up new avenues of scientific research based on data discovery, exploration, fusion, and statistical analysis.   
 
The time domain is the emerging field of astronomical research, as recognized in the 2010 National Research Councils Decadal Survey of Astronomy and Astrophysics \cite{decadal}. Planned facilities for the next decade and beyond, such as the Large Synoptic Survey Telescope (LSST; \cite{lsst}) and the Square Kilometer Array (SKA; \cite{ska}), will revolutionize our understanding of the universe with nightly searches of large swathes of sky for changing objects and networks of robotic telescopes ready to follow up in greater detail anything of interest.

In this paper, we will review the specific data challenges that time domain astronomy presents and the approaches that are being developed to meet them. Much of this work lies at the interface of applied computer science, information technology and astronomy, an area not recognized within the three traditional paradigms of scientific discovery. It is, however, common to contemporary data intensive science and reflected in the development of the field of astroinformatics to address it.

\section{Astronomy in the time domain: real-time mining of massive data streams} 

Just as the rise of information technology enabled the modern panoramic digital sky surveys in the 1990's, the trend has continued with the advent of synoptic sky surveys that cover large areas of the sky repeatedly, looking for moving and variable objects.  An example of such a survey is the Catalina Real-Time Transient Survey (CRTS; htp://crts.caltech.edu; \cite{ref5}, \cite{ref6}, \cite{ref7}).  In a way, we moved from a panoramic digital photography of the sky to a panoramic digital cinematography.  This opening of the time domain for exploration is currently one of the most active and growing areas of research in astronomy, touching upon essentially every part of that science, from the solar system to cosmology, and from stellar evolution to extreme relativistic phenomena \cite{ref8}.  This is a very rich area for scientific exploration and discovery.  Many interesting phenomena, e.g., supernovae and other types of cosmic explosions, can be studied only in the time domain.  While the synoptic sky surveys have resulted in an obvious growth of data volumes, moving astronomy from the Terascale to the Petascale regime, they magnified the already existing challenges in data handling and exploration, and added new ones, e.g., \cite{ref9}, \cite{ref10}.

Probably the most interesting aspect of this is the need for a (near) real-time mining of massive data streams.  Since many of the observed phenomena in this domain are short-lived, and since the scientific returns depend strongly not only on their detection, but also on the timely and well-chosen follow-up observations, there is a need to fully process the data as they stream from the telescopes, compare it with the previous images of the same parts of the sky, automatically and reliably detect any changes, and classify and prioritize the detected events for the rapid follow-up observations.  Analogous situations now exist in many other areas, where the data come continuously from some instruments or sensor networks, and where anomalous or specifically targeted events have to be found and responded to in a rapid fashion.  This poses significant new technological challenges.

\subsection{Data rates}
In the new era of data intensive astronomy, the data rates and volumes are too high for substantial human involvement. Most of the astronomy infrastructure will be automated with robotic telescopes taking observations, workflows in the cloud reducing data and intelligent agent systems evaluating the available information and making and implementing decisions about the next best course of action to further scientific discovery. To get a more quantitative handle on the data rates we are talking about, a useful comparison measure is the Large Hadron Collider (LHC) at CERN. At peak capacity (when all four experiments are running simultaneously), this produces $\sim1.8$ GB/s and requires the largest distributed computing network in the world to handle its output. Since the network can transfer data at $\sim1$ GB/s, we use this as a fiducial value, denoted as 1 LHC.

Table~\ref{table1} gives the data rates for a number of high throughput surveys in terms of LHCs. As can be seen, LSST with event production rates of between $\sim10^5$ and $\sim10^7$ notifications per night is on a par with LHC but the real issue is with the new breed of radio surveys. In primary operation mode, SKA will produce 200 - 2000 PB of data/day (LHC produces $\sim4$ PB/year) -- the scientific equivalent is the detection of 1 core-collapse supernova (massive stellar progenitor) per second over the whole sky originating somewhere in the redshift range $0 < z < 5$ -- and there are operating modes with even higher data rates. Associating and relating these data to themselves and to other data will increase their volume and complexity and is the real challenge facing astronomy.

These extreme surveys define the upper limits for the computational requirements of astronomy in the next twenty years or so. Extrapolating current disk space growth rates to 2030 will put the entire LSST catalog ($\sim200$ TB) easily onto a single disk with plenty of room for associated data. However, conventional relational database technology will almost certainly not scale comparably. Jim Gray stated that RDBMSs do not function well beyond $\sim100$ TB in size and so alternate solutions, such as the NoSQL class of distributed storage technologies for structured data, will be necessary for any of the larger surveys. However, a better match for scientific data is SciDB \cite{scidb}, which is a column-oriented system (rather than row-oriented like a RDBMS) that uses arrays as first-class objects rather than tables but is still ACID. Similarly the processing requirements for LSST are manageable but for SKA, they require a facility within the top 10 of the Top 500 computers on the planet with a few Exaflop capability and software that can run on up to 1 billion cores \cite{skacomp}.

\begin{table}
\caption{Survey data rates in terms of the data rate of the Large Hadron Collider (1 LHC = 1GB/s)}
\label{table1} 
\begin{tabular}{llll}
\hline\noalign{\smallskip}
Survey & Wavelength & Operational & Data rate (LHCs) \\
\noalign{\smallskip}\hline\noalign{\smallskip}
LSST & Optical & 2018 & 0.3 \\
ASKAP & Radio & 2014 & 2 \\
LOFAR & Radio & 2013 & 50 - 200 \\
SKA & Radio & 2020 & 2500 - 25000 \\
\noalign{\smallskip}\hline
\end{tabular}
\end{table}

\section{The Virtual Observatory}

Data intensive astronomy is not only concerned with single monolithic data sets but also with federating disparate data to gain potentially new scientific insights. An obvious example is combining observations of the same objects from different wavelength regimes, e.g., X-ray, infrared, and radio, to understand the various physical processes that contribute to their spectral energy distributions. The time domain adds an extra dimension to this, allowing the identification of temporally correlated behavior, for example, a X-ray burst followed by an infrared burst may indicate the propagation of a shock front from an originating source to circumscribing material.

\begin{table}
\caption{Different types of data access protocol defined by the IVOA.}
\label{table2} 
\begin{tabular}{ll}
\hline\noalign{\smallskip}
Name & Description \\
\noalign{\smallskip}\hline\noalign{\smallskip}
Simple Cone Search (SCS) & Retrieve all objects within a circular region on the sky \\
Simple Image Access (SIA) & Retrieve all images of objects within a region on the sky \\
Simple Spectral Access (SSA) & Retrieve all spectra of objects within a region on the sky \\
Simple Line Access (SLA) & Retrieve spectral line data \\
Simulations (SIMDAL) & Retrieve simulation data \\
Table Access (TAP) & Retrieve tabular data \\
\noalign{\smallskip}\hline
\end{tabular}
\end{table}

The construction of such aggregate data sets from heterogeneous data sources is facilitated by a common set of data access protocols across data archives, common data formats, and common data models. This creates a framework for defining shared elements across data and metadata collections and describing relationships between them so that different representations can interoperate in a transparent manner. The 
International Virtual Observatory Alliance (IVOA; \cite{ivoa}) is in the process of specifying many of these infrastructure components -- Table~\ref{table2} gives a summary of the types of data access protocol defined, many of which have an associated data model and employ VOTable \cite{votable} to serialize tabular data. Other more general data models exist for spatial and temporal metadata, physical units, and observations. VOEvent (defined in section 4.1) may be regarded as the data model describing transient astronomical events and the TimeSeries data model, an extension of a more generalized model for spectrophotometric sequences, is intended to describe any observed or derived quantity that may vary with time.

A secondary tier of IVOA activity deals with components that are more relevant for large numbers of aggregate data sets, including: VOSpace \cite{vospace}, a lightweight common interface to distributed storage; VOPipe, a way to define large-scale data streams; and the IVOA Thesaurus, for describing how common elements are actually defined and relate to each other rather than just identifying and referencing them.

\section{Event infrastructure} 
There is an ever-increasing network of detectors, both ground-based and space-borne, systematically monitoring the sky for changes in either electromagnetic flux (optical, radio, gamma-ray) or something more exotic, such as high energy neutrinos, cosmic rays or gravitational waves. When a significant variation is detected (the significance is determined by such factors as the size, suddenness and duration of the variation as well as the type of detector used), an {\em event notification} is broadcast to all interested parties. This triggers a cascade of activity where the event is placed in context with related data and information: followup observations of the same astronomical event (if possible), associations with other previous or simultaneous observations, both in image and catalog format, analyses and review of such data by both human and machine. This {\em data portfolio} for the event represents a summation of all that is known and understood about it. The most interesting or exciting events will be associated with rich portfolios containing a wide range of heterogeneous material whereas a commonplace event might have a portfolio containing only the original event notification. A portfolio is also a dynamic entity with the potential for new material to be added at any time, from milliseconds to years or even decades after the initial event.

\subsection{VOEvent}
VOEvent \cite{voevent} is a specification for astronomical event notification defined by the IVOA. It defines a data model (and an XML serialization) which captures the minimal semantic set of information necessary to fully describe an event; it address the who, what, why, where, when, and how of the observation. It is interesting to note that Babylonian cuneiform tablets reporting astronomical observations record the same set of information. An infrastructure to support the dissemination of events is also defined with five roles "for the interchange of VOEvent semantics":

\begin{itemize}
\item {\em Author}: the creator of the scientific content of an alert
\item {\em Publisher}: the distributor of an alert
\item {\em Repository}: a persistent store of alerts
\item {\em Subscriber}: a receiver of alerts
\item {\em Broker/Relay}: offers application-level functionality around alerts
\end{itemize}

\begin{figure}
\includegraphics[width=4.5in]{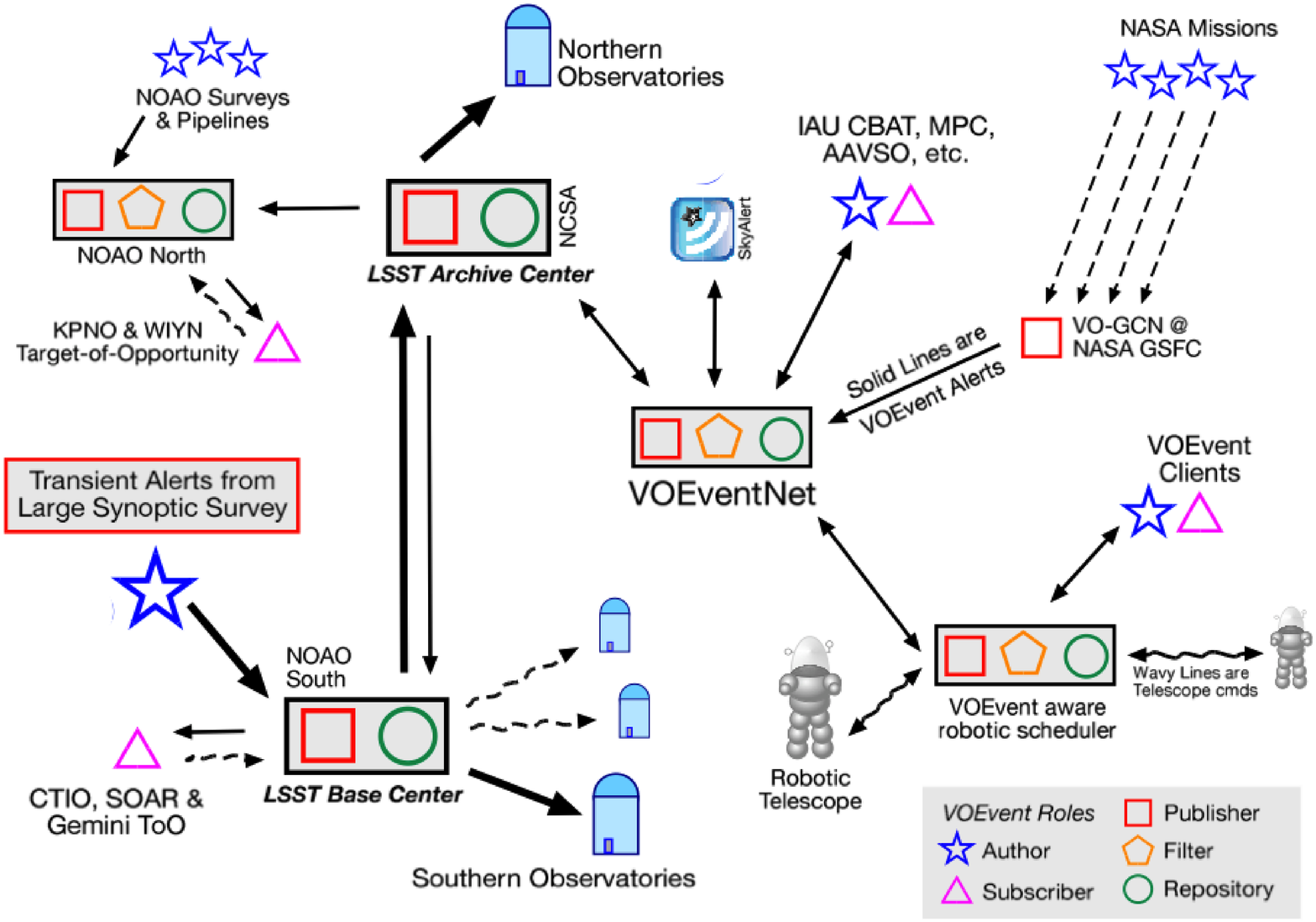}
\caption{This shows a possible architecture for the dissemination of LSST events (courtesy: R. Seaman, NOAO).}
\label{fig1}       % Give a unique label
\end{figure}

This architecture has been proven with the Palomar-Quest \cite{pq} and CRTS \cite{ref5} surveys and is scalable to the new big surveys, e.g., Fig.~\ref{fig1} shows how LSST events could be disseminated.
VOEvent is designed to be transport agnostic so that the most appropriate transport protocol can be used -- currently XMPP/Jabber and a simple TCP-based protocol are used. The metadata descriptions of VOEvent infrastructure components are registered in a distributed resource metadata store, allowing the discovery of {\em event streams} (from particular projects, instruments, etc.) with consistent sets of metadata, as well as servers publishing these with details on how to subscribe to them (which transport protocols are supported), and repositories where particular sets of events are persisted. 

\subsection{SkyAlert service}
Much of the activity of a broker deals with portfolios rather than event notifications -- each VOEvent packet seeds a data portfolio. SkyAlert \cite{skyalert} is a prototype brokering service which allows users to define the types of astrophysical events they are interested in and what happens when a notification of such an event is received by the broker. An {\em alert} is a set of {\em rules} constructed against a particular stream definition. Each rule consists of a {\em trigger} Ð a Pythonic logic expression involving the stream parameters Ð and an optional {\em action} that is enacted when the trigger is true. Portfolios form the input to triggers and actions. An action might be something simple, such as sending an email notification, or a more involved operation, an {\em annotator}, such as fetching archival images or analyzing a light curve \cite{anno}, which generates further events and adds to the portfolio. Events from annotators can therefore cause further actions to be triggered but the SkyAlert system ensures that infinite loops of actions and annotators are avoided. 

\subsection{Event and source crossmatching}
One of the most common annotation activities is to cross-identify events with other data archives, i.e., search for plausible spatial associations between an event and other observations, typically at other wavelengths. The simplest matching criterion is just to take the nearest positional hit but this is not necessarily the best match. Positional accuracies can vary widely between surveys, particularly between different wavelength regimes, leading to multiple possible crossmatch candidates, e.g., a brighter object might have a smaller positional error due to its stronger detection whereas a fainter object might be farther away yet still as likely due to its larger positional error. Other information may also make certain matches far more likely, such as a potential supernova being more likely associated with a nearby galaxy than a star. Several formalisms have been proposed to deal with the general problem of spatial crossmatching but Budavari \cite{budavari} uses Bayesian hypothesis testing to evaluate the quality of candidate associations specifically for detections in space {\em and} time thus allowing inclusion of information about the temporal behavior of particular sources.

A related activity is constructing the time histories of astronomical objects from sets of individual observations of them within the same survey. Varying conditions between observations -- sky brightness, atmospheric, instrumental, etc. -- mean that the same detection thresholds and positional errors cannot be assumed across a survey, e.g., perfect conditions may mean that two nearby sources are resolved on one night but appear as a blended source on another poorer quality night. Constructing the (full transitive) set of associations for $n$ sources from a set of $m$ observations scales at best as ${\cal O}(\sim\!nm^2)$, assuming only one match between individual sets. The PQDR1 catalog of $\sim$10 million sources with typically $\sim$15 observations per source \cite{pqdr1} resulted in a set of over 4 billion associations; with $\sim$500 million sources, each with typically $\sim$200 observations per source, CRTS would have at least $\sim$20 trillion associations. The next generation surveys will take us into the quadrillions and beyond.

Spatial indexing schemes, which provide a single identifier for a region of sky, have been defined, such as HTM \cite{htm} and Healpix \cite{healpix}, and whilst these can speed up individual object lookups in catalogs, there are more efficient ways of doing bulk crossmatches. The Zones algorithm \cite{zones} developed for the SDSS and 2MASS surveys uses a B-tree to bucket two-dimensional space giving dynamically computed bounding boxes (B-tree ranges) for spatial queries. In practice, using an optimal zoning gives several factors of ten increased performance over using indexing schemes, although, in tests with PQDR1, the Quad Tree Cube scheme \cite{q3c} has also shown itself to be equivalently fast.

\section{Classification} 
As the data and event discovery rates increase dramatically, from $\sim$0.1 TB and $\sim$10 -- 10$^2$ events per night now, to $\sim$30 TB and 10$^5$ -- 10$^7$ events per night in the LSST era, available followup facilities would be simply overwhelmed, and unable to react to all potentially interesting events. The essential enabling technologies will be automated, robust classification and decision making for the optimal use of followup facilities. Given the exponential growth of data rates, the traditional ÒmanualÓ approach from the past will simply not scale to the next generation of surveys, especially if one is interested in the rarer transients. The key challenges are thus in the dynamical, real-time characterization and classification of transient events, and the optimal decision making for their followup, and we elaborate on some aspects of that below.

\subsection{Artifacts}
\begin{figure}
\includegraphics[width=4.5in]{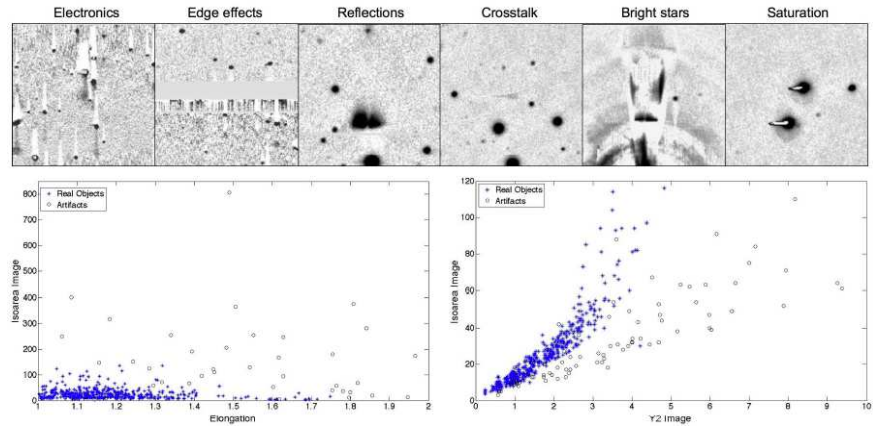}
\caption{This shows an artificial neural network-based artifact classifier for the Palomar-Quest survey \cite{ann}. The top images show the different classes of artifact that the ANN could distinguish. The bottom plots show a couple of morphological parameters used to train the ANN, for which artifacts (+) separate well from genuine objects ($\circ$).}
\label{fig2}       % Give a unique label
\end{figure}

Machine learning has been successfully used with morphological parameters for astronomical image classification. We deployed an artificial neural network-based (ANN) system based on a set of Multilayer Perceptrons to separate real transient sources from a variety of data artifacts \cite{ann} (electronic glitches, saturation, crosstalk, reflections, etc.), as part of the PQ survey's real time data reduction pipeline (see Fig.~\ref{fig2}). Each potential artifact candidate is associated with up to four detections, one for each photometric filter used, but each detection is fed separately to the ANN classifier with a set of morphological parameters obtained from the detection used as input features. The output from the classifier is the probability that an object (detection) is real. While this is a very specialized instance of an automated event classifier for a particular sky survey experiment, it illustrates the plausibility and the potential of this concept. Despite the relatively low number of training cases for many kinds of artifacts, the overall artifact classification rate was around 90\%, with no genuine transients misclassified during our real-time scans.

\subsection{Decision trees}
We are exploring a variety of machine learning techniques to classify transient events using the ongoing CRTS sky survey as a testbed \cite{ref10}. A set of over 60 periodic and non-periodic features are extracted from the light curve of each object of interest (e.g., see \cite{richards}). These features are then used to build a set of decision trees which are able to discriminate between different classes. To reduce the dimensionality of the input space, we have applied a forward feature selection strategy that consists in selecting a subset of features from the training set that best predict the test data by sequentially selecting features until there is no improvement in prediction. Each tree is built using the Gini diversity index as a criterion for choosing the split; the splitting stops when there is no further gain that can be made, and, to avoid overfitting, we use a 10-fold cross validation approach. We find that we can distinguish between blazars, cataclysmic variables and RR Lyrae periodic stars with $>$90\% completeness and $<$10\% contamination, and similarly between different types of supernovae.

\subsection{Symbolic regression}
An automated method of identifying significant multifeature correlations in data offers a different approach to classification. Eureqa \cite{eureqa} is a software tool which aims to describe a data set by identifying the simplest mathematical formulae which could describe the underlying mechanism that produced the data. It employs symbolic regression to search the space of mathematical expressions to determine the best-fitting functional form -- this involves fitting both the form of the equation and its parameters simultaneously. Binary classification can be cast as a problem amenable to this tool -- the ``trick'' is to formulate the search relationship as: $class = g(f(x_1, x_2, ..., x_n))$, where $g$ is the Heaviside step function or the logistic function, which gives a better search gradient. Eureqa finds a best-fit function, $f$, balancing accuracy against complexity, to the data that will get mapped to a 0 or a 1, depending on whether it is positively or negatively valued (or lies on either side of a specified threshold, say 0.5, in the case of the logistic function).  

We have applied this technique to the same data sets ($\sim$60 features) as we did decision trees and find comparable results for completeness ($>$90\%) and contamination ($<$10\%) with 10-fold cross validation when distinguishing between RR Lyrae periodic stars and eclipsing binaries and blazers and cataclysmic variables. However, between Type Ia supernovae (which have a white dwarf progenitor) and core-collapse supernovae (which have a massive stellar progenitor), there is much greater contamination with the latter class, which is probably due to the lower signal-to-noise and sparsity of these data. 

An obvious advantage of this technique is that it provides an analytical expression to separate classes rather than relying on application of a trained black box algorithm as other methods do. This is not only faster, making it more suitable for classification scenarios where speed is important but also, as the least complex solution is favored, identifies those features which may be physically relevant to the classification problem in hand. For example, the skew is one of the three features used in the best-fit function between RR Lyrae and eclipsing binaries and it has been shown \cite{eyer} to be important in semi-automated searches for eclipsing binaries.

\subsection{Bayesian networks}
Bayesian methodology is desirable and attractive for classification work since it can deal with missing data. A Bayesian Network (BN) is a probabilistic graphical model represented through directed acyclic graphs, whose nodes represent variables, and the missing arcs represent conditional independence assumptions. These networks can be used to compute the probability distribution of a subset of variables when other variables are observed (the so-called probabilistic inference). To describe a BN, we need to specify the graph topology and the parameters of each conditional probability distribution and it is possible to learn both from the data. In the Bayesian approach, we generate (and subsequently, update) a library of prior distributions capturing, for example, brightness changes in a certain filter over a certain time interval, conditional on object type such as Type Ia supernova. Such distributions need to be estimated for each type of variable astrophysical phenomenon that we want to classify. An estimated probability of a new event belonging to any given class can then be evaluated.

As a simple demonstration of the technique, we have been experimenting with a prototype na\"ive BN model. We use a small but homogeneous data set involving colors of transients (difference between magnitudes in different photometric filters) detected in the CRTS survey, as measured at the Palomar 1.5-m telescope. We have used multinomial nodes (discrete bins) for 3 colors, with provision for missing values, and a multinomial node for Galactic latitude, which is always present and is a probabilistic indicator of whether an object is galactic or not - those far from the Galactic Plane (high latitude) are more likely to be extragalactic sources. The current priors used are for six distinct classes, which all exhibit single epoch or aperiodic flux outbursts: cataclysmic variables (CVs), supernovae (SNe), blazars (active galaxies with extreme relativistic jets), other types of active galactic nuclei (AGN), UV Ceti stars and everything else bundled into a sixth class.  

With a set of colors from a single epoch as input features, we can distinguish between CVs and SNe with $\sim$80\% completeness and $\sim$19\% contamination, which reflects a qualitative color difference between these two types of transient. Between CVs and blazars, we get $\sim$70 -- 90\% completeness and $\sim$10 -- 24\% contamination, reflecting that colors of these two types of transients tend to be similar, and that some additional discriminative parameter is needed. Eventually we will use a BN with an order of magnitude more classes, more parameters, and additional layers. Measurements from multiple epochs should also improve the classifications.  The end result will be the posteriors for the ``class'' node from the marginalized probabilities of all available inputs for a given object.

\subsection{Probabilistic structure functions}
There are a number of characterizing statistics based on the differences of all possible pairs of data points in a time series, e.g, the discrete correlation function \cite{ek}. Structure functions \cite{sf} typically measure the mean (or some variant thereof) magnitude difference within a particular time-lag range and are used in preference over spectral methods such as power spectrum analysis as the time sampling of astronomical time series is often relatively sparse. However, information is lost in the use of aggregate measures and we have found using the more general 2-dimensional distributions of magnitude changes for different time baselines for all possible epoch pairs in the data set to be a promising discriminator, particularly when they can be optimally binned, e.g., via Bayesian blocks \cite{scargle}. These 2-dimensional ($\Delta m$, $\Delta t$) histograms can be viewed as probabilistic structure functions for the light curves of different types. 

Template distributions for different kinds of transients and variables are constructed using the reliably-classified data with the same survey cadences, S/N, etc. For any newly detected variable or a transient, corresponding ($\Delta m$, $\Delta t$) histograms are accumulated as the new data arrive, and a variety of metrics used to compute the effective probabilistic distances from different templates. The tests so far indicate that classification accuracies in excess of 90\% may be possible using this approach. Generalizations to include triplets or even higher order sets of data points for multi-dimensional histograms are planned.

\subsection{Combining classifiers}
\begin{figure}
\includegraphics[width=4.5in]{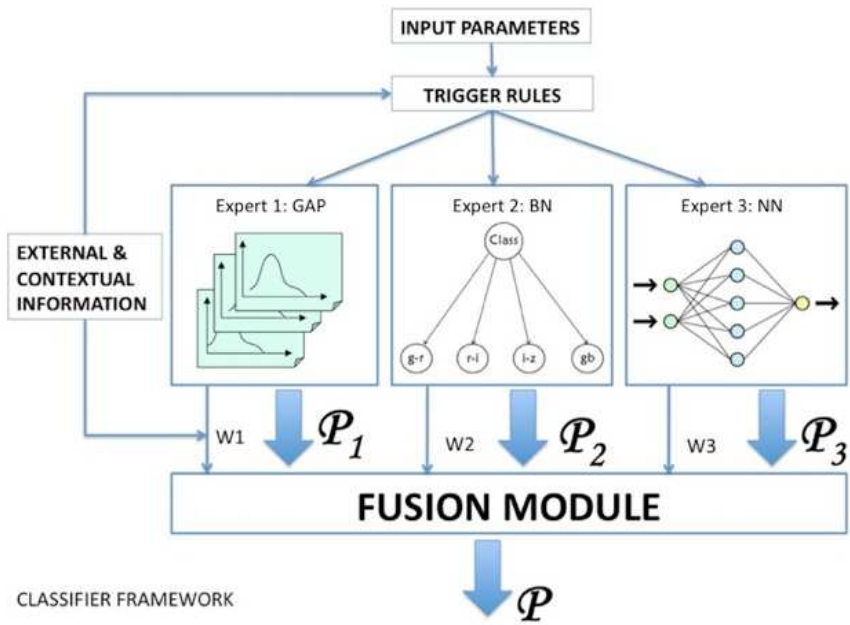}
\caption{A schematic view showing how the results of different classifiers can be combined to obtain an optimal classification.}
\label{fusion}       % Give a unique label
\end{figure}

Given the heterogeneity of data and classes, it would be hard, if not unfeasible, to find a single classifier that suits all needs. The best approach is to use different machine learning models, playing those to their strength. Having a framework that combines results from these ÒindependentÓ classifiers can lead to better overall classification, narrowing down the number of competing classes and thus leading to optimal followups and new discoveries. For example, some classifiers could work better than others in recognizing some classes when certain input parameters are present while some others may be activated only in the presence of certain inputs, e.g., models that cannot deal with missing data. In this context, a sleeping experts framework (see Fig.~\ref{fusion}) can be used: each classifier makes a prediction only when the instance to be predicted falls within its area of expertise. Sleeping experts can be seen as a generalization of the IF-THEN rule: IF the condition is satisfied THEN activate this expert. External and contextual knowledge could be highly relevant to resolving competing interpretations and used to wake up or put experts to sleep and modify online the weight associated with each classifier. Such information, however, can be characterized by high uncertainty and a rich structure. It has been shown that using such a framework is a powerful way to decompose complex classification problems.

As an example, consider such an ensemble of three classifiers, say, a decision tree (DT), a Bayesian network (BN) and a probabilistic structure function (PSF), dealing with a transient alert from a supernova. The light curve for this will normally show very little prior to the initial detection of the event, unless a progenitor or host galaxy was bright enough to be detected, which in this case we will assume it was not. Characterizing statistics from the light curve will carry minimal information and so therefore the DT classifier, which uses them as input features, will either not be triggered or minimal weight will be attached to its results. A PSF for the event will also have little information and, similarly, not be triggered or only with minimal significance attached to its results. For the BN, however, we have colors, proximity information (let us assume a faint nearby spiral), etc. which guarantees its triggering and strong weight attached to its results. In this case, the presence of contextual information was enough to obtain a strong classification, even in the absence of quantitative features, and the sleeping expert framework ensured that the right combination of classifiers was used.

\section{Conclusions}
The field of time domain astronomy illustrates many of the issues common to the emerging data-intensive sciences, particularly those involving sensor networks. Significant detections trigger further automated data gathering or generating actions, building up a body of related information that needs to be collated, managed and, ultimately, appraised and analyzed to determine the optimal response. The techniques and methodologies underpinning these activities belong to a new mode of scientific discovery which recognizes data as the primary focus: experiments exist to generate data which is then searched for scientifically significant patterns rather than to support or test particular hypotheses, with the same data set potentially reused multiple times in different contexts. This change of emphasis places data-related skills, e.g., databases and data mining, at the core, coupled with domain knowledge in the relevant scientific areas. There is thus a challenge to educational institutes to ensure that the next generation of scientists are properly trained to take full advantage of the new data-intensive world.

%\section{Section title}
%\label{sec:1}
%Text with citations \cite{RefB} and \cite{RefJ}.
%\subsection{Subsection title}
%\label{sec:2}
%as required. Don't forget to give each section
%and subsection a unique label (see Sect.~\ref{sec:1}).
%\paragraph{Paragraph headings} Use paragraph headings as needed.
%\begin{equation}
%\end{equation}

%\begin{figure}
%  \includegraphics{example.eps}
%\end{figure}
%\begin{figure*}
 % \includegraphics[width=0.75\textwidth]{example.eps}
%\end{figure*}
%\begin{table}
%\begin{tabular}{lll}
%\hline\noalign{\smallskip}
%\end{tabular}
%\end{table}

\begin{acknowledgements}
This work has been supported in part by the National Science Foundation grants AST-0407448, CNS-0540369, AST-0834235, AST-0909182 and IIS-1118041; the National Aeronautics and Space Administration grant 08-AISR08-0085; and by the Ajax and Fishbein Family Foundations.  We are thankful to numerous colleagues in the VO and Astroinformatics community, and to the members of the DPOSS, PQ, and CRTS survey teams, for many useful discussions and interactions through the years. We thank the anonymous referees for their useful comments.
\end{acknowledgements}



\end{document}